\documentclass[article]{revtex4}
\usepackage{graphicx}
\usepackage{dcolumn}
\usepackage{bm}

\begin{document}

\title{Analog information processing at the quantum limit with a Josephson
ring modulator}
\author{N. Bergeal$^1$, R. Vijay$^{1,2}$, V. E. Manucharyan$^1$, I. Siddiqi$%
^2$, R. J. Schoelkopf$^1$, S. M. Girvin$^1$, M. H. Devoret$^1$}
\affiliation{$^1$Department of Physics and Applied Physics, Yale University, New Haven,
Connecticut, 06520-8284 USA.}
\affiliation{$^2$Department of Physics, University of California, Berkeley CA 94720-7300,
USA.}
\date{\today}

\maketitle

\large
Amplifiers are crucial in every experiment carrying out a very sensitive
measurement. However, they always degrade the information by adding noise.
Quantum mechanics puts a limit on how small this degradation can be.
Theoretically, the minimum noise energy added by a phase preserving
amplifier to the signal it processes amounts at least to half a photon at
the signal frequency ($\frac{1}{2}\hbar \omega _{S}$). In this article, we
show that we can build a practical microwave device that fulfills the
minimal requirements to reach the quantum limit. This is of importance for
the readout of solid state qubits, and more generally, for the measurement
of very weak signals in various areas of science. We also discuss how this
device can be the basic building block for a variety of practical
applications such as amplification, noiseless frequency conversion, dynamic
cooling and production of entangled signal pairs.\newline

The concept of quantum limited amplification was introduced in the 1950's
with the development of the first maser amplifiers\cite{shimoda}. Later,
following the work of Haus and Mullen\cite{haus}, Caves\cite{caves} reviewed
the subject and introduced a general formalism which includes all linear
amplifiers (i. e. an amplifier whose output signal is linearly related to
its input signal). His analysis led to a fundamental theorem : a phase
preserving amplifier has to add a minimum amount of noise to the signal it
processes. The limit is commonly expressed in terms of the minimal
temperature of the noise added by an amplifier to the signal 
\begin{equation}
T_{N}^{\mathrm{added}}=\frac{1}{2}\frac{\hbar \omega _{S}}{k_{B}} 
\label{Caves_Thm}
\end{equation}

where $\omega _{S}$ is the angular frequency of the signal. This corresponds to $\frac{1}{2}$ photon added to each signal mode at the
input of the amplifier. On the other
hand, a phase sensitive amplifier, in which one quadrature is amplified
while the other is de-amplified, is submitted to only a lower limit on the
product of the noise added to the two quadratures and can squeeze the
quantum noise on one quadrature at the expense of extra noise on the other
one. Although such amplifiers can look rather appealing because of their
ability to operate potentially below the quantum limit (\ref{Caves_Thm}),
they are awkward to use in a wide number of applications where both the
phase and amplitude of the signal carry the information. On the other hand,
the so-called non-degenerate parametric amplifier, i. e. an amplifier in
which a non-linear system\cite{louisellbook,tien} with two resonant
frequencies is pumped with an oscillatory source has repeatedly been
suggested to be a good candidate as a phase preserving amplifier reaching
the quantum limit\cite{louisell,gordon}. It operates with two spatially
distinct modes, conventionally called the \textquotedblleft
signal\textquotedblright\ at frequency $\omega _{S}$ and the
\textquotedblleft idler\textquotedblright\ (a.k.a. the "image") at frequency 
$\omega _{I}$. These two modes are coupled in the non-linear system via the
\textquotedblleft pump\textquotedblright\ at frequency $\Omega $. The device
operates as an amplifier with photon number gain when $\Omega =\omega
_{S}+\omega _{I}$ or as a frequency converter without photon number gain
when $\Omega =\left\vert \omega _{S}-\omega _{I}\right\vert $ \cite%
{louisellbook,tien}.

In this article, we will focus on the phase preserving case and show that a
practical, non-degenerate parametric amplifier operating in the microwave
domain can be realized with a simple circuit involving Josephson tunnel
junctions. Because it is minimal in the number of active modes, it should
reach the quantum limit. Unlike microwave SQUIDs, which are powered by a DC
current bias and operate with incoherent Josephson radiation\cite{andre,
spietz}, Josephson parametric amplifiers involve a coherent microwave source.
Josephson tunnel junction parametric amplifiers have so far mainly
focused on degenerate amplifiers ($\omega _{S}=\omega _{I}$) which operate
as phase sensitive amplifiers\cite{yurke88,yurke89,movshovich} and very
little work has been devoted to phase preserving amplifiers. The difficulty
of building a practical device matching the theoretical proposals as well as
the lack of applications requiring quantum limited performances contributed
to put this field on hold. However, recent progress in quantum information
processing using microwave interrogation of solid states qubits\cite%
{wallraff,lupascu,sillamaa, majer} gave rise to a growing need for low-noise
amplifiers which are sensitive enough to measure the extremely weak signals
involved in these new devices and renewed the interest in parametric
amplifiers \cite{tholen,lehnert,castellanos,yamamoto}. Also, amplifiers
operating near the quantum limit are essential in quantum feedback for
sustaining the coherent oscillation of a qubit \cite{korotkov1,korotkov2}.\\

\textbf{\ The Josephson ring modulator}\newline

Even at zero temperature, internal dissipation in a device inevitably adds
noise to the output signals. Thus, it is important to build a completely
dissipationless circuit employing only dispersive elements. The amplifier
that we describe here is based on a particularly interesting novel
non-linear device, which we call the Josephson ring modulator, by analogy
with the ring modulator employing Schottky diodes\cite{Pozar}. The device
consists of four nominally identical Josephson junctions forming a ring
threaded by a magnetic flux $\Phi $. This magnetic flux induces a fixed,
circulating current around the ring. When operated with a bias current lower
than their critical current $I_{0}$, Josephson junctions behave as pure
non-linear inductors with inductance $L_{J}=\varphi _{0}/(I_{0}\cos \delta )$
where $\delta $ is the gauge invariant phase of the junction and $\varphi
_{0}=\frac{\hbar }{2e}$ is the reduced flux quantum. They are the only known
non-linear and non-dissipative circuit elements working at microwave
frequencies. The ring has three orthogonal electrical modes coupled to the
junctions : two differential ones, $X$ and $Y$, and a common one, $Z$ (Fig
1a). They provide the minimum number of modes for 3-wave mixing.\newline

We introduce the node flux $\Phi _{i=1,..,4}$ defined by 
\begin{eqnarray}
V_{i=1,..,4}=\frac{d\Phi _{i=1,..,4}}{dt}
\end{eqnarray}
where $V_{i=1,..,4}$ are the potentials at ring nodes 1, 2, 3 and 4. 
The amplitudes of the three modes $X$, $Y$, $Z$ can be chosen as the following combination of nodes fluxes 

\[
\Phi_{X}=\Phi_{1}-\Phi_{2}\quad ; \quad\Phi_{Y}=\Phi_{4}-\Phi_{3}\quad;\quad
\Phi_{Z}=\Phi_{1}+\Phi_{2}-\Phi_{3}-\Phi_{4} 
\]

In the case of large area junctions, the charging energy due the intrinsic
capacitance of the junctions can be neglected. Hence, the Hamiltonian of the
ring is only given by the sum of the Josephson Hamiltonian of each junction $
H_{J}=-E_{J}\cos \delta _{i=a,b,c,d}$, where $E_{J}=I_{0}\varphi _{0}$ \cite{devoret}. By rewriting the sum of the Josephson energies as a function of
the variables $\Phi _{X}$, $\Phi _{Y}$ and $\Phi _{Z}$ 
\begin{equation}
H_{\mathrm{ring}}=-4E_{J}[\cos \frac{\Phi _{X}}{2\varphi _{0}}\cos \frac{%
\Phi _{Y}}{2\varphi _{0}}\cos \frac{\Phi _{Z}}{2\varphi _{0}}\cos \frac{\Phi 
}{4\varphi _{0}}+\sin \frac{\Phi _{X}}{2\varphi _{0}}\sin \frac{\Phi _{Y}}{%
2\varphi _{0}}\sin \frac{\Phi _{Z}}{2\varphi _{0}}\sin \frac{\Phi }{4\varphi
_{0}}]  \label{energy}
\end{equation}

In figure 1b we plot the energy of local equilibrium states of the Josephson
ring modulator as a function of the magnetic flux $\Phi $ when no external
currents are applied to the ring. There are 4 stable states satisfying the
quantization of the flux through the loop. Although each state is 4$\Phi
_{0} $ periodic as a function of the flux, the envelope of the lowest energy
state remains $\Phi _{0}$ periodic as required by gauge invariance ($\Phi
_{0}=2\pi \varphi _{0}$).\newline

Let us now consider the degenerate ground state at $\Phi =\Phi _{0}/2$
labeled $a$ in figure 1b. For mode intensities $\Phi _{X}$, $\Phi _{Y}$ and $%
\Phi _{Z}$ much smaller than $\Phi _{0}$, we can neglect terms of order
higher than three and (\ref{energy}) reduces to 
\begin{eqnarray}
H_{\mathrm{ring}}=\lambda \Phi _{X}\Phi _{Y}\Phi _{Z}+\mu \lbrack \Phi
_{X}^{2}+\Phi _{Y}^{2}+\Phi _{Z}^{2}] 
\label{energy2}
\end{eqnarray}

with $\lambda =-2\sqrt{2}\pi ^{3}E_{J}/\Phi _{0}^{3}$ and $\mu =\sqrt{2}\pi
^{2}E_{J}/\Phi _{0}^{2}$. Apart from the sought-after pure non-linear
coupling term $\Phi _{X}\Phi _{Y}\Phi _{Z}$, the Hamiltonian contains a
contamination term which is only quadratic in the fluxes and which therefore
only renormalizes the mode frequencies. This powerful result shows that the
Josephson ring modulator can perform the operation of mixing 3 orthogonal
field modes while producing a minimal number of spurious non-linear effects.
The Wheatstone bridge type of symmetry eliminates most of the unwanted terms
in the Hamiltonian, in particular those of the form $\Phi _{X}^{2}\Phi _{Z}$%
, $\Phi _{Y}^{2}\Phi _{Z}$, $\Phi _{X}^{2}\Phi _{Y}$, $\Phi _{Y}^{2}\Phi
_{X} $ which would induce other, unwanted types of mixing ( see below).
Note that
although $\Phi =\Phi _{0}/2$ is optimal for maximizing $\lambda $ while
keeping the working point stable, it is not a stringent condition. In the
following, the differential modes $X$ and $Y$ are used to carry the signal
and the idler with symmetric roles while the $Z$ mode is used as the pump. 
\newline

\textbf{The Josephson Parametric Converter}\newline

\linespread{1.6} We now feed the $X$ and $Y$ modes of the Josephson ring
modulator through two superconducting resonators. A lumped element
representation of the circuit that we have named the Josephson Parametric
Converter (JPC) is shown in figures 2a\&b. The device contains only purely
dispersive elements: superconducting resonators and Josephson junctions.
Since it has no internal dissipation, all the noises appearing at the output
ports originate from the coupling of the JPC to the external circuits
connected at its different ports. There are in fact two possible variations
of the circuit depending on whether the ring modulator junctions are in
parallel with the voltage of the resonators (Fig. 2a) or in series with the
current of the resonators (Fig. 2b). For simplicity and conciseness we treat
here only the first case. The second case can be treated by a simple
extension of the formalism we present. The Hamiltonian of the two resonators
is given by \cite{devoret,yurke84} , 
\begin{eqnarray}
H_{\mathrm{res}}=\frac{{\Phi _{X}^{2}}}{2L_{a}}+\frac{{Q_{X}^{2}}}{2C_{a}}+%
\frac{{\Phi _{Y}^{2}}}{2L_{a}}+\frac{{Q_{Y}^{2}}}{2C_{a}}+H_{\mathrm{damp}}
\end{eqnarray}
where the $\Phi ^{\prime }s$ and the $Q^{\prime }s$ are the conjugated
fluxes and charges in the inductive and capacitive parts of the circuit,
respectively, and where the $L^{\prime }s$ and $C^{\prime }s$ are the
associated inductances and capacitances. The damping term $H_{\mathrm{damp}}$
arising from the coupling to the external source resistors $R_{a}$ and $%
R_{b} $ could be expressed using the Caldeira-Leggett model \cite{caldeira},
which treats dissipation in a quantum circuit, but this detailed description is not of
interest here. In addition, each resonator is submitted to a weak,
time-dependent external drive which models the incoming signal (idler). This
contribution can be taken into account by introducing the Hamiltonian of the
drives \cite{devoret} 
\begin{eqnarray}
H_{\mathrm{drive}}=-\Phi _{X}\frac{U_{1}}{R_{a}}\cos (\omega _{1}t+\phi
_{1})-\Phi _{Y}\frac{U_{2}}{R_{b}}\cos (\omega _{2}t+\phi _{2})
\end{eqnarray}

The pump mode is assumed to be so stiffly driven that $\Phi _{Z}$ can be
regarded as an imposed oscillating classical field, which does not suffer
backaction from its coupling to the other modes.\newline
Therefore, the total Hamiltonian of the JPC is given by 
\begin{eqnarray}
H_{JPC}=H_{\mathrm{res}}+H_{\mathrm{ring}}+H_{\mathrm{drive}}
\end{eqnarray}

In the following, we consider the case where the pump is driven with two
tones at frequencies $\Omega _{\sigma }=\omega _{1}+\omega _{2}$ and $\Omega
_{\delta }=\omega _{1}-\omega _{2}$ (we assume $\omega _{1}>\omega _{2}$)
and corresponding current amplitude $I_{\sigma }^{p}$ and $I_{\delta }^{p}$.
Using the Hamilton equations $\dot{Q}_{X}=-\partial H_{JPC}/\partial \Phi
_{X}$ and $\dot{Q}_{Y}=-\partial H_{JPC}/\partial \Phi _{Y}$, we can derive
the equations of motion for the two modes $X$ and $Y$ 
\begin{eqnarray}
\ddot{\Phi}_{X}+\kappa _{a}\dot{\Phi}_{X}+\omega _{a}^{2}\Phi _{X}+2\Phi
_{Y}[\frac{\chi _{\sigma }}{C_{a}}\cos (\Omega _{\sigma }t+\varphi _{\sigma
})+\frac{\chi _{\delta }}{C_{a}}\cos (\Omega _{\delta }t+\varphi _{\delta
})] &=&2\epsilon _{1}\cos (\omega _{1}t+\phi _{1}) \\
\ddot{\Phi}_{Y}+\kappa _{b}\dot{\Phi}_{Y}+\omega _{b}^{2}\Phi _{Y}+2\Phi
_{X}[\frac{\chi _{\sigma }}{C_{b}}\cos (\Omega _{\sigma }t+\varphi _{\sigma
})+\frac{\chi _{\delta }}{C_{b}}\cos (\Omega _{\delta }t+\varphi _{\delta
})] &=&2\epsilon _{2}\cos (\omega _{2}t+\phi _{2})
\end{eqnarray}%
where $\chi _{\sigma }=I_{\sigma }^{p}/(4\varphi _{0})$ and $\chi _{\delta
}=I_{\delta }^{p}/(4\varphi _{0})$. The coefficients $\kappa _{a(b)}=\left(
R_{a(b)}C_{a(b)}\right) ^{-1}$ are the usual damping factors in RLC
circuits, $\omega _{a(b)}=\sqrt{\frac{L_{a(b)}+L_J}{L_JL_{a(b)}C_{a(b}}}$
are the resonance frequencies of resonators renormalized by the quadratic
 terms in (\ref{energy2}) and $\epsilon _{1(2)}=U_{1(2)}\kappa
_{a(b)}$. Note that the presence of higher order spurious terms would make the resonance frequencies dependent on pump power and induce
instabilities. Our circuit has a direct mechanical analog consisting of two
coupled harmonic oscillators whose mutual coupling is parametrically driven.
Following the usual treatment of a parametric amplifier, we impose the
resonant tuning $\omega _{1}=\omega _{a}$ and $\omega _{2}=\omega _{b}$ and
look for solutions of the form $\Phi _{X}=xe^{i\omega _{1}t}+c.c.$ and $\Phi
_{Y}=ye^{i\omega _{2}t}+c.c.$. Keeping only the terms oscillating at $\omega
_{1}$ and $\omega _{2}$, we obtain the phasors

\begin{eqnarray}
x=\frac{-i\kappa _{b}\omega _{2}\tilde{\epsilon}_{1}-\frac{\tilde{\chi}%
_{\sigma }}{C_{a}}\tilde{\epsilon}_{2}^{\ast }+\frac{\tilde{\chi}_{\delta }}{%
C_{a}}\tilde{\epsilon}_{2}}{\kappa _{a}\kappa _{b}\omega _{1}\omega _{2}-%
\frac{\chi _{\sigma }^{2}}{C_{a}C_{b}}+\frac{\chi _{\delta }^{2}}{C_{a}C_{b}}%
}  \label{eqmotion1}
\end{eqnarray}

\begin{eqnarray}
y=\frac{-i\kappa _{a}\omega _{1}\tilde{\epsilon}_{2}-\frac{\tilde{\chi}%
_{\sigma }}{C_{b}}\tilde{\epsilon}_{1}^{\ast }+\frac{\tilde{\chi}_{\delta
}^{\ast }}{C_{b}}\tilde{\epsilon}_{1}}{\kappa _{a}\kappa _{b}\omega
_{1}\omega _{b}-\frac{\chi _{\sigma }^{2}}{C_{a}C_{b}}+\frac{\chi _{\delta
}^{2}}{C_{a}C_{b}}}  \label{eqmotion2}
\end{eqnarray}

where $\tilde{\chi}_{\delta }=\chi _{\delta }e^{i\varphi _{\delta }}$, $%
\tilde{\chi}_{\sigma }=\chi _{\sigma }e^{i\varphi _{\sigma }}$ and $\tilde{%
\epsilon}_{1}=\epsilon _{1}e^{i\phi _{1}}$, $\tilde{\epsilon}_{2}=\epsilon
_{2}e^{i\phi _{2}}$.

From the point of view of microwave circuits, rather than the local fluxes $
\Phi _{X}$ and $\Phi _{Y}$ and voltages $U_{1}$ and $U_{2}$, it is more
convenient to introduce the normalized amplitudes of the incoming and
outgoing modes $a^{_{\mathrm{in}}}_{1(2)}$ and $a^{_{\mathrm{out}}}_{1(2)}$ at ports 1 and
2. This transformation is described in details in the Methods section. As a
result, we can express the equations (\ref{eqmotion1}) and (\ref
{eqmotion2}) in a very concise way by introducing the scattering matrix $S_{
\mathrm{JPC}}$ of the JPC.

\begin{eqnarray}
\left( 
\begin{array}{c}
a_{1}^{_\mathrm{out}} [\omega_1] \\ 
a_{1}^{\ast_ \mathrm{out}} [-\omega_1] \\ 
a_{2}^{_\mathrm{out}} [\omega_2] \\ 
a_{2}^{\ast_ \mathrm{out}}[-\omega_2]%
\end{array}%
\right) &=&\left( 
\begin{array}{cccc}
r_{1}& 0 & t_{1} & s_{1}\\ 
0 & r_{1}& s^{\ast }_{1} & t^{\ast }_{1} \\ 
t_{2} & s_{2} & r_{2} & 0 \\ 
s^{\ast }_{2} & t^{\ast }_{2} & 0 & r_{2}%
\end{array}%
\right) \cdot \left( 
\begin{array}{c}
a_{1}^{_\mathrm{in}} [\omega_1] \\ 
a_{1}^{\ast_\mathrm{in}} [-\omega_1] \\ 
a_{2}^{_\mathrm{in}} [\omega_2] \\ 
a_{2}^{\ast_\mathrm{in}}[-\omega_2]%
\end{array}%
\right)
\end{eqnarray}

where the coefficients are given, at the resonant tuning, by $r_{1}=r_{2}=r=\frac{1-|\rho _{\delta }|^{2}+|\rho
_{\sigma }|^{2}}{1+|\rho _{\delta }|^{2}-|\rho _{\sigma }|^{2}}$, $t_{1}=t_{2}^{\ast }=t=\frac{%
2i\rho _{\delta }}{1+|\rho _{\delta }|^{2}-|\rho _{\sigma }|^{2}}$ and $s_{2}=s_{1}=s=%
\frac{-2i\rho _{\sigma }}{1-|\rho _{\delta }|^{2}+|\rho _{\sigma }|^{2}}$
and where we have introduced the reduced pump currents $\rho _{\delta }=\frac{\tilde{\chi}_{\delta }}{%
\sqrt{C_{a}C_{b}\kappa _{a}\kappa _{b}\omega _{1}\omega _{2}}}$ and $\rho
_{\sigma }=\frac{\tilde{\chi}_{\sigma }}{\sqrt{C_{a}C_{b}\kappa _{a}\kappa
_{b}\omega _{1}\omega _{2}}}$. The three coefficients $r$, $t$ and $s$
satisfy the relation $|r|^{2}+|t|^{2}-|s|^{2}=1$. The form of this
scattering matrix is in fact quite remarkable. As we show in the Methods
section, $S_{\mathrm{JPC}}$ has the exact minimal form required to perform
phase preserving amplification with minimum added noise and noiseless
frequency conversion. This is the consequence of (i) the dispersive nature
of the operation of the device and (ii) the number of modes having been kept
minimal. The same matrix form is obtained with the series circuit of Fig.
2b, albeit with different expressions for the $\rho ^{\prime }s$.

The case $\rho _{\delta }=0$ corresponds to the optimal amplification
operation described in Fig 2c. The coefficients $|r|$ and $|s|$ can then be
written as $|r|=\sqrt{G}=\frac{1+|\rho
_{\sigma }|^{2}}{1-|\rho _{\sigma }|^{2}}$ and $|s|=\sqrt{G-1}=
\frac{2\rho _{\sigma }}{1-|\rho _{\sigma }|^{2}}$. Amplification ($G\gg1$) is
obtained when the reduced pump current $\rho _{\sigma}$ approaches $1$ from
below. The diagonal term $r$ can be seen as a photon \textquotedblleft
cis-gain\textquotedblright\ characteristic of 1-port reflection amplifier
operation. From the point of view of each port separately, the device
behaves as a sort of ideal negative resistance: the incoming wave at either
port is reflected with a power gain $G$ and its phase is preserved, when the
signal at the other port is zero. A circulator is needed to separate the
outgoing wave from the incoming one. The non-diagonal term $s$ can be seen
as a photon \textquotedblleft trans-gain\textquotedblright\ between
different ports. Since it couples conjugated mode amplitudes $a^{_{\mathrm{
out}}}_{1(2)}$ and $a^{\ast _{\mathrm{in}}}_{2(1)} $, the device behaves as a phase
conjugating frequency converter with power gain $G-1$ \cite{louisell}. In
particular, this operation can be used to mix down a signal from high
frequency $\omega _{1}$ to low frequency $\omega _{2}$. The remarkable
feature here is the presence of photon gain. SIS mixers operating from the
quasiparticle branch of a tunnel junction are so far the only known
practical examples of mixers with gain, and they can operate quite close to
the quantum limit \cite{mears}.

The case $\rho _{\sigma }=0$ corresponds to a conversion mode where an
incoming mode at one port is partially reflected and partially converted
into the second mode (Fig 2d). This operation is analogous to the one
performed by a beam splitter but with the peculiarity that the frequency of
the transmitted signal is converted when modes 1 and 2 have different
frequencies $\omega _{1}$ and $\omega _{2}$. The device conserves the total
number of incoming photons ($|r|^{2}+|t|^{2}=1$), whereas the energy is
conserved only if $\omega_{1}$= $\omega _{2}$. Pure frequency conversion
with unity gain can be obtained when $|\rho _{\delta }|=1$. Although both
modes of operation make frequency conversion possible, there are some
fundamental differences between the two processes. The pure converter case
allows to convert frequency with no added noise and without any reflection.
On the other hand, the phase conjugating conversion of the amplification
mode has the advantage of enabling photon gain. \newline

\textbf{\ Noise of the JPC}\newline

Let us now analyze the noise properties of the JPC for the two different
cases of operation. Assuming thermal equilibrium with $T\ll \frac{\hbar
\omega }{k_{B}}$, each port is fed at its input with half a photon of noise
arising from vacuum fluctuations. Therefore, the total output noise power
emitted by each port in units of photon number per mode is 
\[
N^{\mathrm{out}}=\frac{1}{2}|r|^{2}+\frac{1}{2}|t|^{2}+\frac{1}{2}|s|^{2}=%
\frac{1}{2}(2(|r|^{2}+|t|^{2})-1) 
\]%
In the case of amplification ($t=0$) with large gain ($|r|\gg $1), the noise
referred back to the input is 
\[
N_{\mathrm{eff}}^{\mathrm{in}}=N^{\mathrm{out}}/|r|^{2}(\simeq N^{\mathrm{out%
}}/|s|^{2})=1+\frac{-1}{2|r|^{2}}\rightarrow 1 
\]%
Although each port is fed at its input with only half photon of noise, after
amplification the total output noise at each port is equivalent to one
photon at the input  ($N_{\mathrm{eff}}^{\mathrm{in}}\rightarrow1$). This is an illustration of Caves' theorem \cite{caves}:
the noise added  by the amplification process to the vacuum noise already
present at the input port is equivalent to half a photon ($N_{\mathrm{add}}^{\mathrm{in}}\rightarrow \frac{1}{2}$).

In the pure converter case ($|s|=|r|=0$, $|t|=1$), the noise referred back
to input is 
\[
N_{\mathrm{eff}}^{\mathrm{in}}=N^{\mathrm{out}}/|t|^{2}=\frac{2|t|^{2}-1}{%
2|t|^{2}}=\frac{1}{2} 
\]%
In this case, the output noise is identical to the input noise (half photon)
and no noise is added during the process ($N_{\mathrm{add}}^{\mathrm{in}}= 0$). The photon number gain is unity,
despite the fact that it is possible to have power gain when the frequency
is up-converted.\newline

\textbf{The general case of arbitrary detuning}\\

Detuning the signal frequencies from the resonance frequencies of the
resonators ($\omega _{1}\neq \omega _{a}$ and $\omega _{2}\neq \omega _{b}$)
complicates the expression of $S_{JPC}$  but
retains the phase-preserving quantum limited operation to be reached when $%
\rho _{\delta }=0$ or $\rho _{\sigma }=0$. In the general case ($r\neq 0$, $%
s\neq 0$ and $t\neq 0$) the matrix does not retain the phase preserving
property. \\
In the amplification mode of operation ($\rho_{\delta =0}$), the coefficients of $S_{
\mathrm{JPC}}$ are given by
\begin{eqnarray}
r_{1,2}=-\frac{(\vartheta _{2,1}+i)(\vartheta _{1,2}+i)-|\rho _{\sigma}|^{2}}{(\vartheta _{2,1}+i)(\vartheta _{1,2}-i)-|\rho _{\sigma }|^{2}} \quad \text{and} \quad s_{1,2}=\frac{-2i\rho _{\sigma }}{(\vartheta _{2,1}+i)(\vartheta
_{1,2}-i)-|\rho _{\sigma }|^{2}}
\label{ex_gain}
\end{eqnarray}

where $\vartheta _{1}=\frac{(\omega _{1}^{2}-\omega _{a}^{2})Q_a}{\omega
_{a}^2}$ and $\vartheta _{2}=\frac{(\omega _{2}^{2}-\omega _{b}^{2})Q_b}{%
\omega _{b}^2}$. Here we have introduced the quality factor of the resonators $Q_{a}=\frac{\omega_a}{\kappa_a}$ and $Q_{b}=\frac{\omega_b}{\kappa_b}$. Figure 3a shows a typical example of gain curves for different values of $|\rho _{\sigma }|$. In the large gain limit the expression of $r_{1,2}$ reduces to a Lorentzian form
\begin{eqnarray}
r_{1,2}\simeq\frac{\sqrt{G}}{\sqrt{1+G\big(\frac{Q_a}{\omega_a}+\frac{Q_b}{\omega_b}\big)^2\big(\omega_{1,2}-\omega_{a,b}\big)^2}}
\end{eqnarray}
The -3dB bandwidth of the amplifier is thus
\begin{eqnarray}
B=\frac{2}{\sqrt{G}}\Big(\frac{Q_a}{\omega_a}+\frac{Q_b}{\omega_b}\Big)^{-1}
\label{ex_b}
\end{eqnarray}

We arrive here at an important result: the bandwidth of the amplifier is inversely proportional to the amplitude gain $\sqrt{G}$. This feature is a general property of parametric amplifiers.   At this point, we would like to stress that the pump frequency $\Omega _{\sigma }$ is an additional tuning parameter, which allows to displace the center of the signal bandwidth within the larger tuning bandwidth of the resonators.  The case of an arbitrary detuning for the conversion mode of operation is treated in the Methods section.\\

\textbf{Practical issues: gain, bandwidth, dynamic range and stability.}%
\newline

We now analyze some practical issues and show that we can build a practical
device which would be useful for many different applications. The questions
of power gain and bandwidth are central and are intimately related.
Ideally, the amplified noise should be much larger than the noise of the following amplifier in the measurement chain. For a
quantum limited amplifier working at GHz frequencies and assuming the best
\textquotedblleft state of the art\textquotedblright\ commercial device as a
following amplifier (a noise temperature of a few K is typical), the power gain has to be at least 20 dB, although of course a
smaller power gain can still lead to an improvement in the overall system
noise temperature. To optimise the ability of the amplifier to follow fast signals,  a bandwidth around the carrier frequency as large as possible is sough. However, as shown by  
relation (\ref{ex_b}), the parametric coupling imposes the signal bandwidth to decrease with the amplitude gain. Although the gain of the JPC should in principle reach any
arbitrarily large value when $|\rho _{\sigma }|$ approaches $1$ sufficiently
close, two limitations can occur.
The first limitation is that when $|\rho _{\sigma }|\rightarrow 1^{-}$, the
fraction of the pump current feeding the junctions should remain well below
the Z mode critical current $I_{0}'=I_{0}\cos \frac{\Phi 
}{4\varphi _{0}}=\frac{I_0}{\sqrt{2}}$, in order for the parametric amplification to
remain stable (higher order non-linear terms invade the behavior of the
device as the critical current is reached). It is useful to rewrite the
expression of $|\rho _{\sigma }|$ as 
\[
|\rho _{\sigma }|=\frac{1}{4}\sqrt{Q_{a}Q_{b}p_{a}p_{b}}\frac{
I_{\sigma }^{p}}{I'_{0}}
\]

in which we introduce the participation ratios of the inductance of the
Josephson ring modulator to the resonators inductance 
$p_{a,b}=\frac{L_{a,b}}{L_J+L_{a,b}}$
in the parallel case and 
$p_{a,b}=\frac{L_J}{L_J+L_{a,b}}$ in the series case. Since each junction receives a fourth of the total pump current, the first limitation
thus translates into\\

\begin{equation}
\sqrt{Q_{a}Q_{b}p_{a}p_{b}}>1 
 \label{1st_limitation}
\end{equation}

Figure 3b shows the constraints on the JPC bandwidth and gain $G$ imposed by this limitation. This figure illustrates the impossibility of obtaining at the same time a high gain value and a large bandwidth with a parametric amplifier. Although this figure would seem to suggest that the participation ratio $p_a$ and $p_b$ should be as high as 
possible, in practice dynamic range considerations limit this possibility (see below).
The second limitation arises from the fact that  the  sum of the resonator energies, each being weighted by its participation ratio, cannot exceed the Josephson energy. We write this new condition as

\begin{equation}
E_{a}p_{a}+E_{b}p_{b}<E_{J}
\end{equation}

In particular, the amplified zero-point quantum noise cannot exceed the Josephson energy
\begin{equation}
G\hbar (p_{a}\omega _{a}+p_{b}\omega _{b})/2<E_{J}
\label{lim2}
\end{equation}

Taking $p_{a}=p_{b}=p$ to simplify the algebra we can rewrite Eq. (\ref{lim2}) as 

\begin{equation}
G<\frac{Z_{Q}}{Z_{c}}p^{0,-2}  \label{Gain_limitation}
\end{equation}

where
$Z_{Q}=\frac{\varphi _{0}^{2}}{\hbar }=\frac{\hbar }{\left( 2e\right) ^{2}} \simeq 1\mathrm{k}\Omega$ is the quantum of impedance and $Z_{c}=\frac{\omega _{a}+\omega _{b}}{\sqrt{C_{a}C_{b}}\omega _{a}\omega _{b}}$ is an impedance characterising the resonators. Using a conventional microwave technology, this impedance would be of the order of 50$\Omega$.The exponents zero and -2 refer to the parallel and series case respectively.

The power gain $\times$ bandwidth product is an important characteristics which determines the total flow of information that can be processed by the amplifier.  Equations  (\ref{1st_limitation}) and (\ref{Gain_limitation}) can be combined to obtain an important bound on this product

\begin{equation}
G\times B=\frac{2\omega }{Q}G^{1/2}<2\omega \sqrt{\frac{Z_{Q}}{Z_{c}}}p^{+1,0}
\end{equation}
where we have taken $\omega_a$=$\omega_b$=$\omega$ to simplify the algebra.
Since $p<1$, the final upper bound on the gain$\times $bandwidth product is
thus $G\times B <2\omega \sqrt{\frac{Z_{Q}}{Z_{c}}}$. Thus, both parallel and perpendicular circuits have the same limitation on the power gain $\times $bandwidth product. However,  in the case of  the parallel circuit the maximum gain is strongly constrained by relation ($\ref{Gain_limitation}$). Therefore, the series case appears more favourable in most of the practical cases. 
Another important characteristic of an amplifier is its dynamic range, i. e., for a given gain, the maximum input power $P_{\max}$ that the device can amplify before it starts to saturate. The same considerations involving the maximum power produced by the device, as developed in relation ($\ref{lim2}$),  can be used to obtain the dynamic range:

\begin{equation}
2Gp\Big(\frac{P_{\max }}{B}+\frac{\hbar\omega}{2}\Big)<E_{J } 
\end{equation}

Therefore

\begin{equation}
P_{\max }<\frac{B}{2}\Big(\frac{E_J}{Gp}-\hbar\omega\Big)
\end{equation}

However,  when the input power becomes too large, our small amplitude
approximation is no longer valid and higher non-linear terms in (6) start to
play a role. Therefore, experimentally,  the amplifier may  saturate before reaching the theoretical value.\\

We now turn to the question of stability. The point $\rho _{\sigma }=1$
corresponds to the onset of spontaneous self-oscillations of the system.
Therefore, the JPC should be operated at a distance from this critical point
safe in regards to fluctuations in pump drive power. But here, the situation
is better controlled than in previous studies where optimization of gain
would conflict with an increase of noise caused by the proximity of a
poorly identified instability \cite{wahlstein, mygind, bryant} whose influence might be difficult to avoid.\newline

\textbf{\ Production of entangled signal pairs and dynamic cooling}\newline

The gain of the JPC is high enough to potentially raise the level of quantum
fluctuations at a much higher than the level of the noise of the second amplifier
in the chain. An interesting experiment consists in turning on the pump
without feeding any signals at the input ports of the JPC.
Quantum-mechanically, the pump can still produce output signals, which can
be seen as arising from the amplification of zero-point motion fluctuations.
Moreover, since the scattering matrix conserves the volume of the phase
space, the amplified noise appearing at the two ports must be entirely
correlated. The function which is performed is two mode squeezing\cite{bjork}%
. As the output signals can have many real microwave photons, such a device
could be used for analog quantum encryption\cite{grangier}.

Another interesting feature of the pure frequency converter mode of
operation is that, unlike the amplifier mode, it has no added noise at the
output. The JPC device operating in this case with a unity photon number
``trans-gain'' can swap the photons at the two ports and be used as a
refrigerator. Suppose that the frequency at port 2 is much smaller than the
frequency at port 1, which sees an environment cold in the sense $%
\hbar\omega_{1}\gg k_{B}T_{1}$. Initially port 2 is seeing an environment
which is hot in the sense $\hbar\omega_{2}\ll k_{B}T_{2}$. When the JPC is
operated, the photons at port 2 are shuttled to port 1 where they are
evacuated, while zero-point photons from port 1 go in the other direction to
replace the photons at port 2 imposing vanishing temperature. The cooling
rate being $\dot{q}=k_{B}T_{2}\kappa_{b}$, the refrigeration power is only
of the order of 1pW at 4K and for a bandwidth of 1GHz, but it can be very
useful for a high-Q resonator isolated from the thermal bath.\newline

\textbf{\ Conclusion}\newline

The Josephson Parametric Converter would fill a niche which was up to now
unavailable in the landscape of microwave processing devices, that of
3-wave mixing for non-degenerate parametric amplification operating at the
quantum limit. Moreover, we would like to stress that the present level of
control in the dynamics of tunnel junctions in resonant circuits, as
demonstrated by recent several successful operations \cite%
{siddiqi,wallraff,metcalf, lupascu, sillamaa}, ensures that its realization
is entirely within reach. This development would bring the subject of analog
rf quantum signal processing (should we nickname it quantum radioelectricity ?) 
to a qualitatively new level.\newline

Correspondence and requests for material should be address to M. D and N.B

This work was supported by NSA through ARO Grant No. W911NF-05-01-0365, the
Keck foundation, and the NSF through Grant No. DMR-032-5580. M. H. D.
acknowledges partial support from College de France.\newline

\textbf{\ Methods}\newline
\large
\textbf{\ Transformation of local fluxes and voltages to traveling waves}%
\newline

In the circuit of figure 2a, the local fluxes and voltages can be expressed
as a function of the amplitude of the incoming and outgoing modes $A^{in}$
and $A^{out}$ at ports 1 and 2 using the following relations 
\[
A_{1}^{_\mathrm{in}}+A_{1}^{_\mathrm{out}}=\frac{V_{1}-V_{2}}{\sqrt{R_{a}}}=%
\frac{i\omega _{1}x}{\sqrt{R_{a}}};\quad A_{1}^{_\mathrm{in}}+A_{1}^{_%
\mathrm{out}}=\frac{V_{4}-V_{3}}{\sqrt{R_{b}}}=\frac{i\omega _{2}y}{\sqrt{%
R_{b}}};\quad A_{1}^{_\mathrm{in}}=\frac{U_{1}}{2\sqrt{R_{a}}}:\quad A_{2}^{_%
\mathrm{in}}=\frac{U_{2}}{2\sqrt{R_{b}}} 
\]%
$A^{_\mathrm{in}}$ and $A^{_\mathrm{out}}$ are expressed in square root of
watts. We can now define the normalized amplitude $a^{_\mathrm{in}}$ and $%
a^{_\mathrm{out}}$ expressed in square root of photon number per unit time. 
\[
a_{1}^{_\mathrm{in}}=\frac{A_{1}^{_\mathrm{in}}}{\sqrt{\hbar \omega _{1}}}%
;\quad a_{2}^{_\mathrm{in}}=\frac{A_{2}^{_\mathrm{in}}}{\sqrt{\hbar \omega
_{2}}};a_{1}^{_\mathrm{out}}=\frac{A_{1}^{_\mathrm{out}}}{\sqrt{\hbar \omega
_{1}}};\quad a_{2}^{_\mathrm{out}}=\frac{A_{2}^{_\mathrm{out}}}{\sqrt{\hbar
\omega _{2}}} 
\]

At this point, the normalized ampitudes are still classical variables. The
passage to the creation and annihilation operators is performed by the
simple replacement $a\rightarrow \hat{a}$ and $a^{\ast }\rightarrow \hat{a}%
^{\dag }$.\\

\textbf{\ Minimal scattering matrix for quantum information processing}%
\newline

In order to perform information processing at the quantum limit, a device
must fulfill requirements that impose constraint of its scattering matrix $S$%
. In this section we derive the minimal form of $S$ to perform phase
preserving amplification with minimal added noise and noiseless frequency
conversion in the case of a device involving only two modes. Following a route similar but not identical to that pioneered by Caves, we
introduced the generalized scattering matrix $S$ of a linear microwave
device which relates input and output modes at its different ports.

\[
\Lambda ^{out}=S\cdot \Lambda ^{in} 
\]%
Here we have introduced the mode amplitude input and output vectors 
\[
\Lambda ^{in}=\left( 
\begin{array}{cccc}
a_{1}^{_\mathrm{in}} &  &  &  \\ 
a_{1}^{\ast _\mathrm{in}} &  &  &  \\ 
\vdots &  &  &  \\ 
a_{n}^{_\mathrm{in}} &  &  &  \\ 
a_{n}^{\ast_\mathrm{in}} &  &  &  \\ 
&  &  & 
\end{array}%
\right) \quad ;\quad \Lambda ^{out}=\left( 
\begin{array}{cccc}
a_{1}^{{_\mathrm{out}}} &  &  &  \\ 
a_{1}^{\ast _\mathrm{in}} &  &  &  \\ 
\vdots &  &  &  \\ 
a_{n}^{_\mathrm{in}} &  &  &  \\ 
a_{n}^{\ast_\mathrm{in}} &  &  &  \\ 
&  &  & 
\end{array}%
\right) 
\]%
where the symbol $^{\ast }$ denotes the complex conjugation. The $a_{n}$ are
normalized mode amplitudes expressed in square root of photon number per
unit time. Although they are at first treated as classical scalar fields,
they can, in a later quantum mechanical treatment, be formally replaced by
annihilation ($a_{n}\rightarrow \hat{a}_{n}$) and creation ($a_{n}^{\ast
}\rightarrow \hat{a}_{n}^{\dagger }$) operators. Both $a$ and $a^{\ast }$
have to be present in the input and output vectors because of possible phase
conjugating processes coupling an $a^{_{out}}$ to an $a^{\ast _{in}}$.
In the case of a device with two ports, the most general matrix has only 8
independent complex coefficients.

\[
S= \left( 
\begin{array}{cccc}
r_{1} & u_{1} & t_{1} & s_{1} \\ 
u^*_{1} & r^*_{1} & s^*_{1} & t^*_{1} \\ 
t_{2} & s_{2} & r_{2} & u_{2} \\ 
s^*_{2} & t^*_{2} & u^*_{2} & r^*_{2}%
\end{array}
\right) 
\]

Our requirement of information processing at the quantum limit implies that
the scattering matrix must describe a canonical transformation that
preserves the commutation relations of the bosonic fields $[\hat{a}_{n}^{_%
\mathrm{out}},\hat{a}_{n}^{\dagger _\mathrm{out}}]=[\hat{a}_{n}^{_\mathrm{in}%
},\hat{a}_{n}^{\dagger_\mathrm{in}}]$. Mathematically, this is translated by
the property of symplecticity of the $S$ matrix \cite{arvind,guillemin}

\[
^TSJS=J 
\]
where 
\[
J= \left( 
\begin{array}{cccc}
0 & 1 & 0 & 0 \\ 
-1 & 0 & 0 & 0 \\ 
0 & 0 & 0 & 1 \\ 
0 & 0 & -1 & 0%
\end{array}
\right) 
\]
In order to perform phase preserving amplification, we need to impose that
the reflection at each port separately preserves the phase of the signal
(i.e. a phase shift of the incoming wave results in an identical phase shift
of the outgoing wave). That implies that $u_{1}=u^*_{1}=u_{2}=u^*_{2}=0 $.
Finally, without loss of generality we can impose the modulus of the
reflection coefficients to be identical at each port ($|r_{1}|=|r_{2}|=|r|$%
). It follows that $S$ has the minimal form 
\[
S= \left( 
\begin{array}{cccc}
|r|e^{i\alpha_1} & 0 & |t|e^{i\beta_1} & |s|e^{i\gamma_1} \\ 
0 & |r|e^{-i\alpha_1} & |s|e^{-i\gamma_1} & |t|e^{-i\beta_1} \\ 
-|t|e^{i(\alpha_1+\alpha_2-\gamma_1)} & |s|e^{i(\alpha_2-\alpha_1+\beta_1)}
& |r|e^{i\alpha_2} & 0 \\ 
|s|e^{-i(\alpha_2-\alpha_1+\beta_1)} & -|t|e^{-i(\alpha_1+\alpha_2-\gamma_1)}
& 0 & |r|e^{-i\alpha_2}%
\end{array}
\right) 
\]
where the coefficients are linked by the relation $|r|^2+|t|^2-|s|^2=1$.

The case $s=0$ corresponds to a conversion operation where an incoming mode
at one port is partially reflected and partially converted into the second
mode. Since $|r|^2+|t|^2=1$, the total number of photons is conserved during
the process. The case $t=0$ corresponds to an amplification operation since
the total number of photons is not longer conserved ($|r|^2+|s|^2\neq1$).
Therefore, $|r|$ can take any value larger than one, the phase of the signal being preserved. This matrix has the
simplest form for performing phase preserving amplification and
frequency conversion at the quantum limit.\newline

\textbf{Case of  arbitrary detuning for the conversion mode of operation }%
\newline
In the conversion mode of operation ($\rho _{\sigma =0}$) 
\begin{eqnarray}
r_{1,2}=\frac{(\vartheta _{2,1}-i)(\vartheta _{1,2}+i)-|\rho _{\delta
}|^{2}}{(\vartheta _{2,1}-i)(\vartheta _{1,2}-i)-|\rho _{\delta }|^{2}}\quad \text{and} \quad t_{1,2}=\frac{-2i\rho _{\delta }}{(\vartheta _{2,1}-i)(\vartheta
_{1,2}-i)-|\rho _{\delta }|^{2}}
\end{eqnarray}

where $\vartheta _{1}=\frac{(\omega _{1}^{2}-\omega _{a}^{2})Q_a}{\omega
_{a}^2}$ and $\vartheta _{2}=\frac{(\omega _{2}^{2}-\omega _{b}^{2})Q_b}{
\omega _{b}^2}$.

\textbf{Fig 1.} Electrical modes and energy states of the Josephson ring
modulator. a) The Josephson ring modulator consists of four nominally identical
Josephson junctions ($a$, $b$, $c$ and $d$) and has four orthogonal
electrical modes. The two differential modes $X$,$Y$ and the common mode $Z$
are coupled to the junctions whereas the 4th mode $W$ remains uncoupled. b)
Energy states of the ring modulator. There are 4 stable states satisfying
the relation $\delta_{a}+\delta_{b}+\delta_{c}+\delta_{d}=n\frac{2\pi\Phi}{
\Phi_{0}}$ where $\delta_{i}$ is the gauge invariant phase of the junction $
i $. Each state is 4$\Phi_{0}$ periodic as a function of the flux $\Phi$
through the loop but the envelope (blue line) of the lowest energy remains $
\Phi_{0}$ periodic. Other energy extremum states are not represented here. At point $a$, the device can produce a pure non-linear
coupling term $\Phi_{X}\Phi_{Y}\Phi_{Z}$ whose contamination is only of the
type $\Phi_{X}^2+\Phi_{Y}^2+\Phi_{Z}^2$. Point $a$ actually corresponds to
two degenerate states separated by an energy barrier whose height is $2(\sqrt{2}-1)E_J$. At point $b$, there is
no contamination and the non-linearity is of the purest form, but it would be very difficult 
 to stabilise the device in an excited state.\newline

\textbf{Fig 2.} Description of the Josephson Parametric Converter. a) Lumped
element schematic of the parallel JPC. The device is based on a ring modulator
coupled to two parallel $LC$ resonators corresponding to the two differential modes $X$
and $Y$. The common mode $Z$ is driven by a current source $I^{p}$. Both
resonators are coupled to external drives. b) Lumped element schematic of
the series JPC. c) Scattering representation of the JPC in the case of
amplification operation. Here the white arrows denote the conjugation
operation since the non diagonal terms of the scattering matrix couples $a^{
\mathrm{out}}$ to $a^{\ast \mathrm{in}}$. d) Scattering representation of
the JPC in the case of the pure conversion operation.\newline

\textbf{Fig 3.}  Gain of the JPC. The
figure displays in color scale the gain $r_1$ in the amplification
mode as a function of the normalised input frequency $\omega _{1}/\omega_a$ for different values
of $|\rho _{\sigma }|$ (from 0.6 to 0.99). In this example, $Q_a=50$ and the damping factors are taken to be identical for the two resonators($\omega_a/Q_a$=$\omega_b/Q_b$).\\

\textbf{Fig 4.} Main constraint on the gain$\times$bandwidth product of the JPC. The figure displays
in color scale the pump current in the junction, normalised by the $Z$  mode critical current, as a function of the relative bandwith $B/\omega_a$ and the gain G. The brown area corresponding is not accessible for the JPC since in this region, the pump current in the junction always exceeds their critical current. The different contours correspond to the various limitations obtained for different participation ratios $p_a$ and $p_b$. 

\begin{figure}[p]
\centering
\includegraphics[width=15cm]{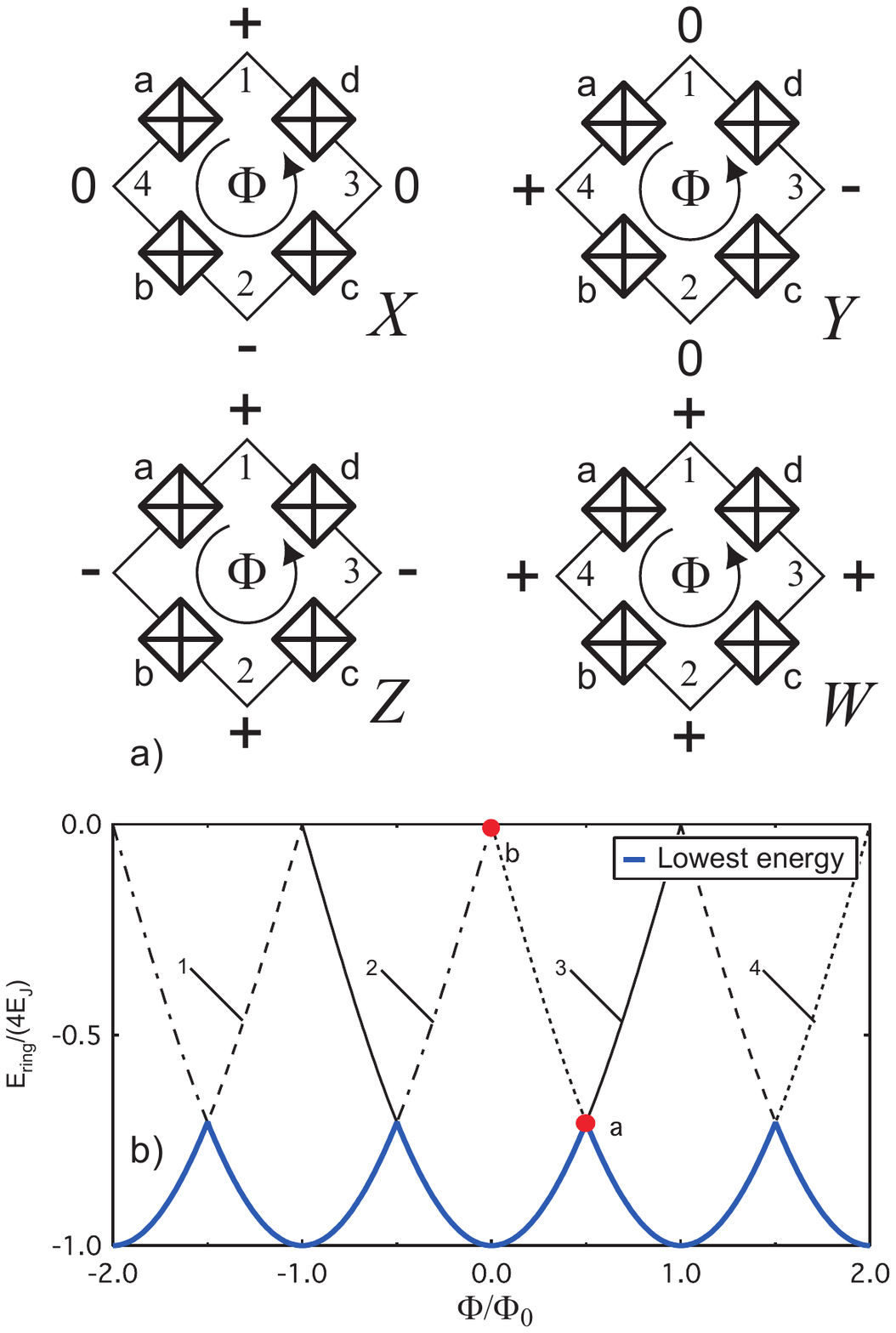}
\caption{}
\end{figure}

\begin{figure}[p]
\centering
\includegraphics[width=15cm]{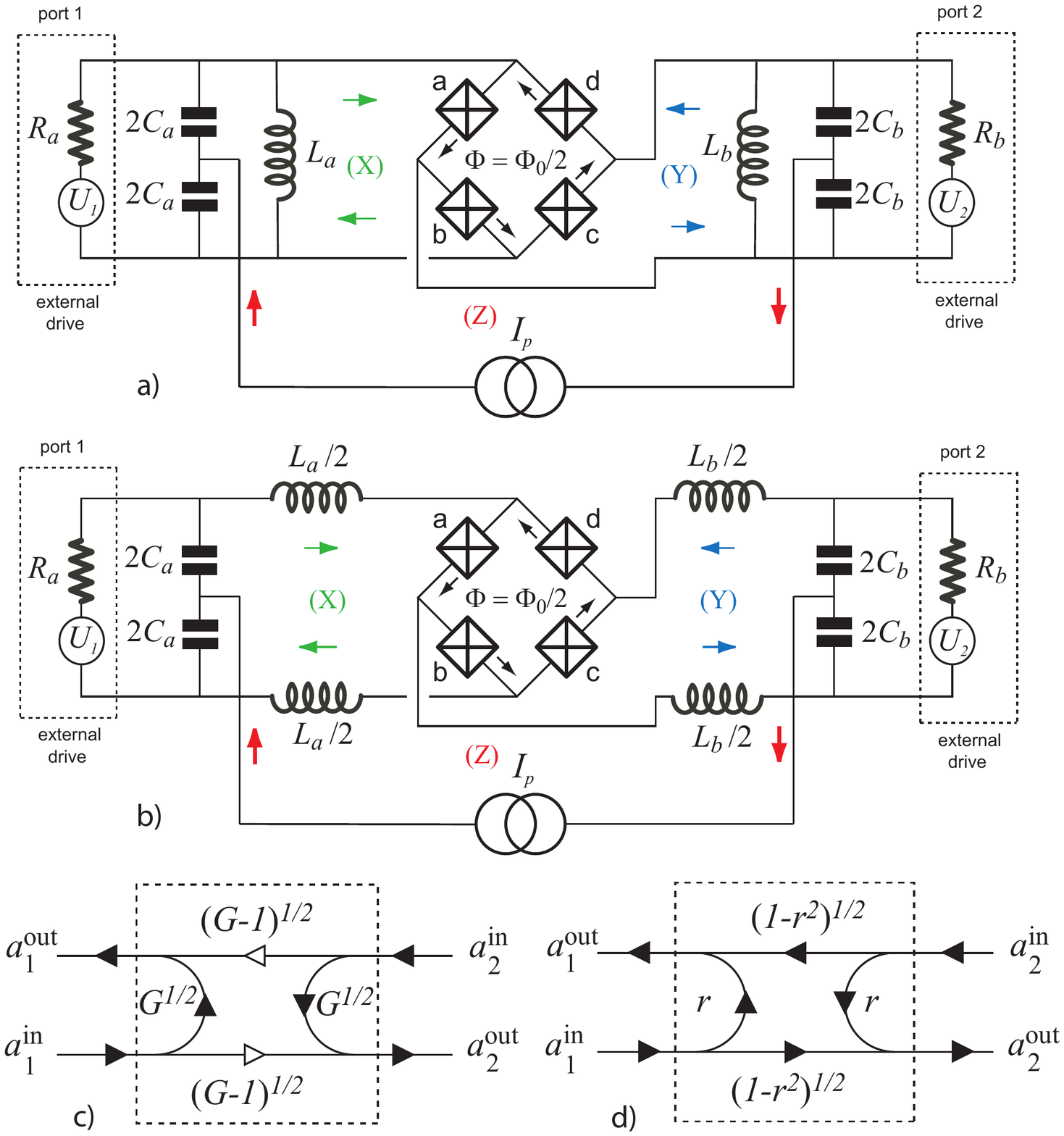}
\caption{}
\end{figure}

\begin{figure}[p]
\centering
\includegraphics[width=15cm]{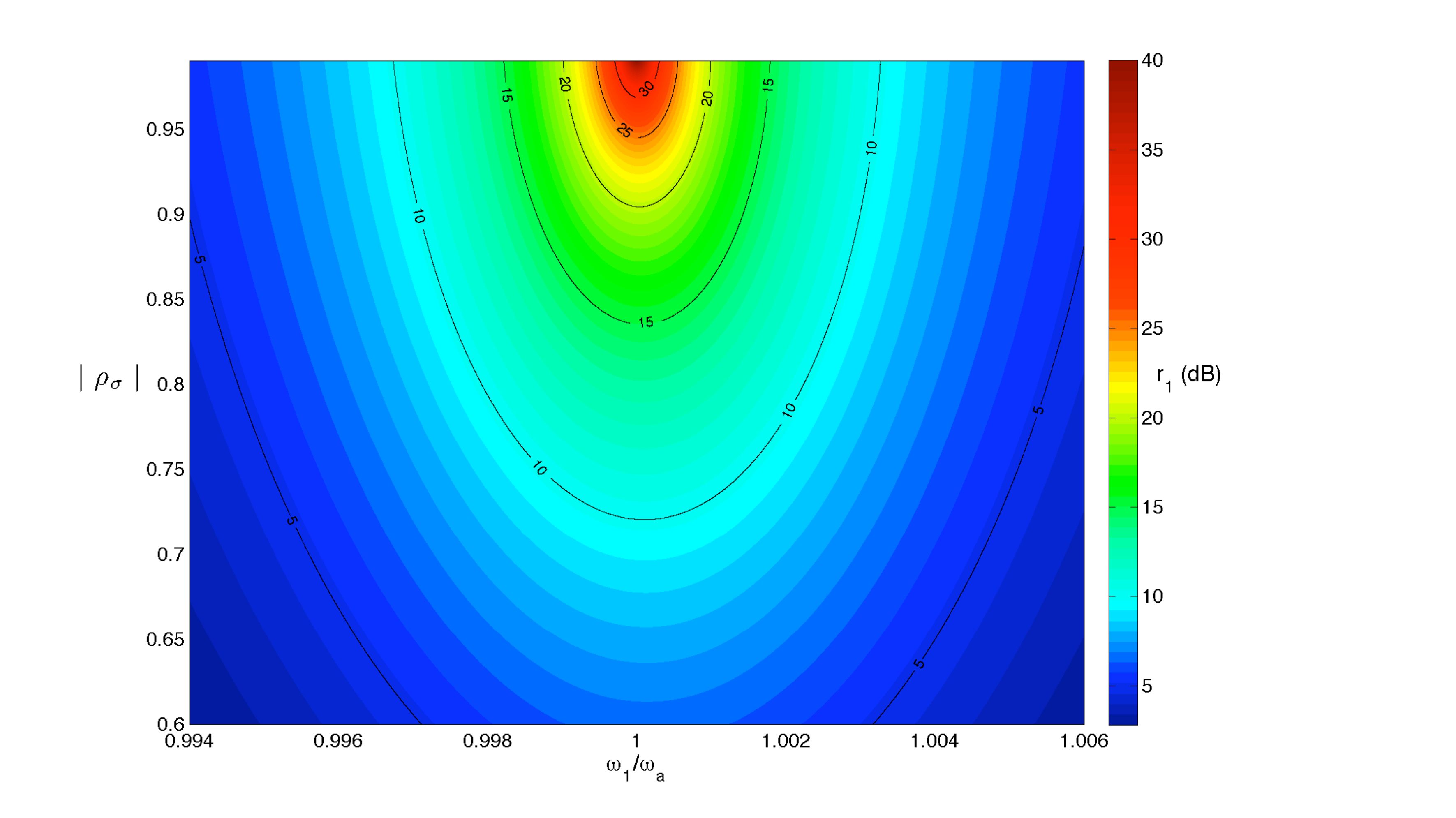}
\caption{}
\end{figure}

\begin{figure}[p]
\centering
\includegraphics[width=15cm]{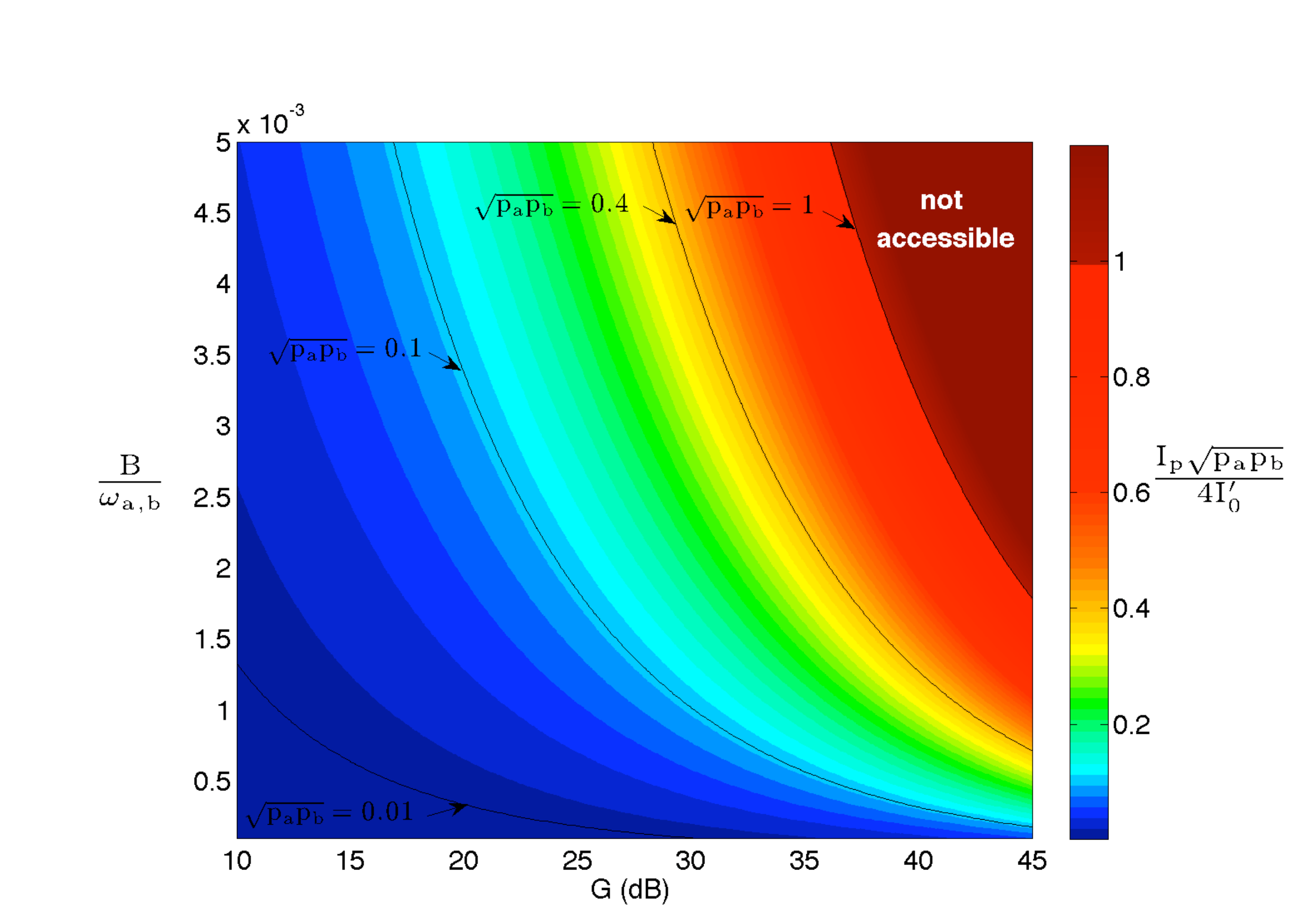}
\caption{}
\end{figure}

\end{document}